\documentstyle[12pt,a4]{article}

\input epsf
\newcommand{\postscript}[2]
{\setlength{\epsfxsize}{#2\hsize}
\centerline{\epsfbox{#1}}}

\def\calo{{\cal O}}
\def\call{{\cal L}}

\def\ie{{\it i.e.}}

\def\be{\begin{equation}}
\def\ee{\end{equation}}
\def\bea{\begin{eqnarray}}
\def\eea{\end{eqnarray}}

\begin{document}
\begin{titlepage}
\hfill CERN--TH/95--142
\vspace{4.0cm}
\center{\large\bf SHOULD SPARTICLE MASSES UNIFY}
\center{\large\bf  AT THE GUT SCALE?\footnote{Talk given at the
30th Rencontres de Moriond on ``Electroweak Interactions and
Unified Theories'', Les Arcs, France, March 11-18, 1995.}
}
\vspace{1.0cm}
\center{{\sc Alex Pomarol}}
\center{{\it Theory Division, CERN}\\{\it CH-1211 Geneva 23,
Switzerland}}
\vspace{4cm}
\begin{abstract}
Gauge  and Yukawa (for the third family)  coupling unification
seem to be  the best predictions of the grand-unified
 theories (GUTs).
In supersymmetric GUTs, one also expects that  the sparticle masses
 unify at the GUT scale (for
 sparticles embedded in the same GUT multiplet).
I show under what circumstances GUTs do not lead to
sparticle mass unification.
In particular, I give examples of  SU(5)
and SO(10) SUSY GUTs
in
which  squarks and sleptons of a family have different tree-level
masses at the
unification scale.
The models have interesting relations between
Yukawa couplings. For example, I present
an SO(10) GUT that  allows for a large ratio
of the top to bottom Yukawas,
accounting for  the large $m_t /m_b$.
The splittings can also be induced in the Higgs soft masses
and    accommodate the electroweak breaking.
\vskip .3in
\noindent CERN--TH/95--142\hfill\\
\noindent May$\,\,$1995\hfill\
\end{abstract}
\end{titlepage}

\section{Introduction}

Several reasons  persuade us to  study
supersymmetric (SUSY) grand-unified theories (GUTs) and their
low-energy predictions.
Two motivations that  I find very  appealing are:
\begin{itemize}
\item Gauge coupling unification: Taking the values of the
three gauge couplings at the weak scale  and extrapolating
 (using the RGEs)
to high energies, one finds that they  meet at a scale $M_G\sim 10^{16}$ GeV
\cite{dg,drw1}.
\item The matter field multiplets, which under the standard model group
look as in arbitrary representations, can be embedded in
few multiplets of the GUT \cite{gg}. For example in SU(5), the matter fields
(of a given generation)
are embedded in the ${\bf \bar 5}$ and  {\bf 10}, and in SO(10) they
surprisingly fit in a single {\bf 16}.
\end{itemize}
The above hints strengthen the hypothesis that
  ($i$) there is a manifest GUT at $M_G\sim
10^{16}$ GeV and ($ii$)  the effective theory below $M_G$ is just the
supersymmetric standard model (MSSM) \cite{dg} (desert hypothesis).
 The desert hypothesis acts as a microscope that
magnifies by 14 orders of magnitude
and allows us to look at Planckean distances;
it
 allows us to translate
theoretical ideas near the Planck mass
 into low-energy predictions.
There are many more parameters in addition to gauge couplings
to be used as probes of SUSY GUTs: fermion and
sparticle masses and mixing angles add up to a grand total of 110
physical
parameters
just in the supersymmetric flavour sector.
According to the desert
hypothesis, each of  these parameters carries direct information about the
structure of the theory at Planckean distances.

Here I will focus on the sparticle masses \cite{us}.
It is often assumed that
the  \underline{sparticles are} \underline{degenerate at $M_G$
 (universality)}; it was
motivated from the need to suppress flavour-violating processes
\cite{dg}. Nevertheless,
 sparticles in
different
multiplets of the unified group
 have no symmetry reason to be degenerate at
$M_G$ and even if they are assumed to be
degenerate at the Planck scale (as in minimal
supergravity theories) their different interactions will
split
them by the time they reach the unification scale \cite{hkr,pre,pp}.
These
splittings are typically large due to the large size of the
representations in unified groups \cite{pre,pp}.

Although this strong form of universality is not very realistic
in GUTs,
it is widely believed that a more restricted
form
is always valid: \underline{sparticles belonging to the
same multiplet} \underline{are
degenerate at the unification scale}.
 This is considered a direct
consequence of unification, which will be experimentally checked if
 sparticles are discovered at the LHC and NLC and their masses are
known with some precision.

I will show  that, because the GUT group is spontaneously broken,
there is no good reason for this belief \cite{us}. Sparticles, such
as $\tilde b_R$ and $\tilde\tau_L$,  which by virtue
of
their gauge and family quantum numbers can be grouped into an
irreducible SU(5) multiplet, do not necessarily originate from {\it one}
such multiplet; they could come about from a linear combination of
several  multiplets. This can produce non-degeneracy among
sparticles of the same generation and complementary SU(5) quantum
numbers\footnote{These are sparticles whose combined quantum numbers
complete an irreducible SU(5) multiplet.}. I will show that
these sparticle
splittings occur precisely in theories (and for the same reasons)
that produce interesting and
desirable relations among fermion masses.
\pagebreak

\section{Non-unified sparticle masses in an SU(5) model}

The model \cite{us} is a minimal extension of the SU(5) SUSY GUT
\cite{dg}. Consider an SU(5) theory with just the
 third generation consisting of a $\bar{\bf 5}_1$ and ${\bf 10}_1$, the
usual
Higgs fiveplet ${\bf H}$ and antifiveplet $\overline{\bf H}$,
 and the adjoint
${\bf
24}$ that breaks SU(5) down to SU(3)$\times$SU(2)$\times$U(1) at the
unification scale $M_G$  by acquiring a vacuum expectation
value (VEV) that points in the
hypercharge direction:
\be
\langle {\bf 24}\rangle =V_{24}{\bf Y}\equiv V_{24}\, {\rm diag}
(2,2,2,-3,-3)\, .
\label{vevsigma}
\ee
The bottom and tau masses are given by the superpotential
\be
W=h\,
{\bf 10}_1 \overline{\bf H}\,  \bar{\bf 5}_1\, ,
\ee
and are equal at the unification scale \cite{ceg}.
Now add an extra fiveplet and antifiveplet denoted by ${\bf 5}$
and
$\bar{\bf 5}_2$    with the following couplings:
\be
W={\bf 5}\left[M\,\bar{\bf  5}_1+\lambda\, {\bf 24}\, \bar{\bf
5}_2\right]
+h\, {\bf 10}_1 \overline{\bf H}\,  \bar{\bf 5}_1\, ,
\label{superpotential}
\ee
where $M$ is near the unification mass. One linear combination of
 $\bar{\bf 5}_1$ and
 $\bar{\bf 5}_2$ will acquire a large mass of order $\sim
M_{G}$. The orthogonal combination will be part of the low-energy
spectrum. It contains the right-handed bottom quark and the tau
lepton
doublet which are denoted by $D^c$ and $L$ respectively;  because the
hypercharges of $D^c$ and $L$
differ, it follows from eqs.~(\ref{vevsigma})
and (\ref{superpotential}) that they will be {\it different}
linear combinations of the corresponding states in  $\bar{\bf 5}_1$
and
$\bar{\bf 5}_2$:
\be
\left(\matrix{D^c\cr L}\right)=-\sin\theta_Y \bar{\bf 5}_1
+\cos\theta_Y \bar{\bf  5}_2\, ,
\label{rotation}
\ee
where
\be
\sin\theta_Y=\frac{\rho{\bf Y}}{\sqrt{1+\rho^2 {\bf Y}^2}}\, ,
\ee
with $\rho=\lambda V_{24}/M$.

Since  $\bar{\bf 5}_1$ and $\bar{\bf 5}_2$ are in different
representations
of  SU(5), they have, in general, different soft SUSY-breaking masses
at $M_G$ \cite{pp}:
\be
\call_{soft}=m^2_1|\bar{\bf 5}_1|^2+m^2_2|\bar{\bf 5}_2|^2\, .
\ee
Since the  light combination is given by (\ref{rotation}), one has
\bea
m^2_{\tilde b_R}&=&m^2_2+ s^2_{b_R}(m^2_1-m^2_2)\, ,\nonumber\\
m^2_{\tilde\tau_L}&=&m^2_2+ s^2_{\tau_L}(m^2_1-m^2_2)\, ,
\label{splitst}
\eea
where $b_R\in D^c$, $\tau_L\in L$ and $s_a$  is given by
\be
s_a=\sin\theta_{Y_a}=\frac{\rho{ Y_a}}{\sqrt{1+\rho^2 {Y_a^2}}}\, ,
\ee
and $Y_a$ is the hypercharge of $a$.
Therefore, the squark and slepton masses differ at $M_G$;
their fractional mass-splitting, for $\Delta>0$, is given by
\be
\frac{m^2_{\tilde\tau_L}-m^2_{\tilde b_R}}{m^2_{\tilde\tau_L}}
=\frac{s^2_{\tau_L}-s^2_{b_R}}{\Delta+s^2_{\tau_L}}\,
,
\label{splits}
\ee
where $\Delta=m^2_2/(m^2_1-m^2_2)$.
Eq.~(\ref{splits}) is plotted in Fig.~1. One can see that
a mass-splitting of
  $\sim 30\%$ can be obtained.

The fermion masses arise from  the Yukawa  coupling
${\bf 10}_1\overline{\bf H}\,\bar{\bf 5}_1$:
\bea
m_b&=&hs_{b_R}\langle\overline{\bf H}\rangle\, ,\nonumber\\
m_\tau&=&hs_{\tau_L}\langle\overline{\bf H}\rangle\, ,
\eea
which leads to the  ratio between the bottom and
tau mass
\be
\frac{m_b}{m_\tau}=\frac{s_{b_R}}{s_{\tau_L}}=
\frac{2}{3}\sqrt{\frac{1+9\rho^2}{1+4\rho^2}}\, .
\label{splitf}
\ee
This ratio  tends to $2/3$ and $1$
in the small and large  $\rho$ limit respectively.
{}From eqs.~(\ref{splits}) and (\ref{splitf}), one can see that the
scalar  mass-splitting is correlated to the fermion one.
This is shown in Fig.~1, where the dashed line represents the ratio
$m_b/m_\tau$. The maximum values for the scalar mass-splitting
correspond to
 $m_b/m_\tau\sim 0.7$--$0.8$.
\begin{figure}[t]
\postscript{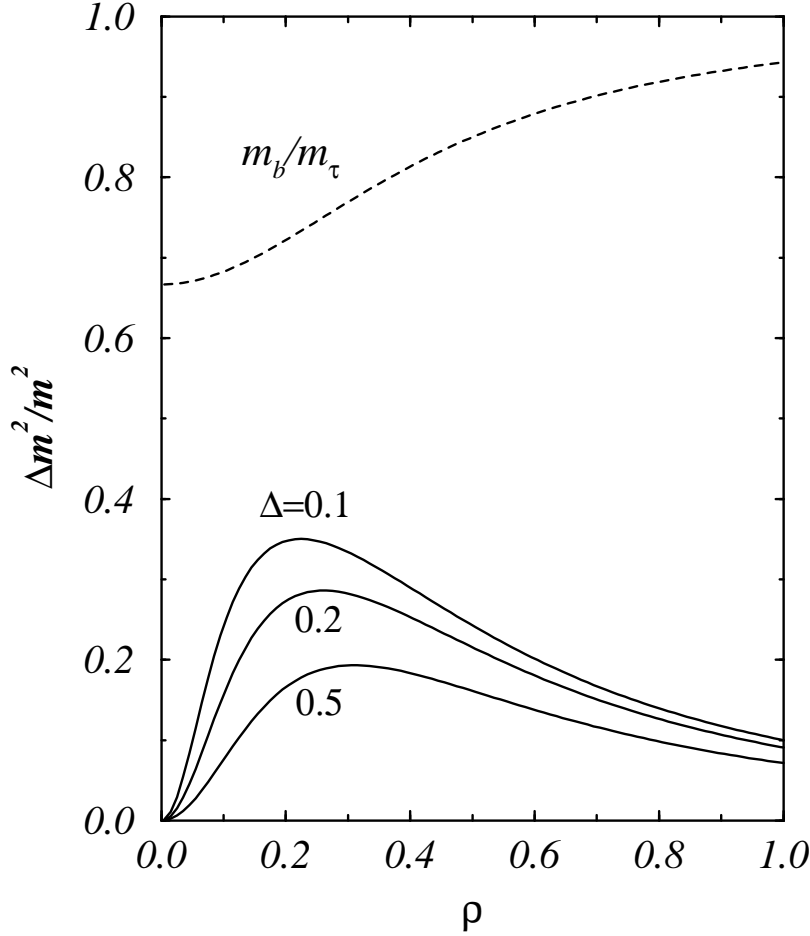}{0.6}
\caption {Scalar mass-splitting $\Delta m^2/m^2\equiv
(m^2_{\tilde\tau_L}-m^2_{\tilde b_R})/m^2_{\tilde\tau_L}$
 as a function of $\rho=\lambda V_{24}/M$ and for
different values of $\Delta=m^2_2/(m^2_1-m^2_2)$. The dashed line
corresponds to the ratio $m_b/m_\tau$.}
\end{figure}
Thus this model, although it is a minimal perturbation of the SU(5)
SUSY GUT \cite{dg}, easily accommodates values of $m_b/m_\tau$
 that are between $2/3$ and
$1$. As a consequence,
 the strong constraints on the top mass
that arise
from
bottom--tau unification \cite{langacker} can be relaxed.

It is now easy to see how the sparticle and particle splittings came
about in this model. Although the right-handed bottom and the tau
lepton doublet -- by virtue of their family and gauge quantum numbers
-- appear to belong to the same $\bar{\bf 5}$ of SU(5), they in fact,
because of their different hypercharges,  came from two {\it
different } linear combinations of a pair of $\bar{\bf 5}$s. This
causes
SU(5)-breaking effects in sparticles and particles
to be felt at the tree-level, since they occur at the very basic
stage of defining the light states of the theory\footnote{
See ref.~\cite{83} for other uses of this mechanism.}.

The same idea can be implemented for the
states in the decuplets, adding to the previous model an extra
 $\overline{\bf 10}$ and  ${\bf 10}_2 $.
 As shown in Fig.~2 of ref.~\cite{us},
 these
splittings
can be much larger than for the fiveplet since
the differences in hypercharges are larger in the decuplet.

\section{An SO(10) model with large $h_t/h_b$}

The previous  example  can be easily adapted to SO(10);
the Higgs are in the ${\bf 45}\ni {\bf 24}$ and
 ${\bf 10}_H\ni \{{\bf H},\, \overline{\bf  H}\}$,
and the matter fields are embedded in the ${\bf 16}$ spinor
representations. Here, however, I will consider a
different scenario.
Instead of adding an extra $\overline{\bf 16}$ and ${\bf 16}$,
I will  add a ${\bf 10}$, ${\bf 10'}$ and a
 ${\bf 16}_H$ that gets a VEV of $\calo(M_G)$:
\be
W={\bf 10'}\left[M {\bf 10}+\lambda\, {\bf 16}_H\,
{\bf 16}\right]+h\,
{\bf 16}{\bf 10}_H\,  {\bf 16}\, .
\label{superpotentialc}
\ee
As in the previous model, the light quarks and leptons
arise from  the linear combination,
 $-\sin\theta{\bf 10}+\cos\theta{\bf 16}$, where now
 the mixing angle is (because  only the SU(5) singlet of ${\bf 16}_H$
gets a VEV)
\be
\sin\theta=\cases{{\rho}/{\sqrt{1+\rho^2}}&
for $D^c$ and $L$,\cr
0& for the rest of the fields,\cr}
\label{sin}
\ee
where $\rho=\lambda\langle {\bf 16}_H\rangle/M$.
Notice that only $D^c$ and $L$ are a mixture of both multiplets, the
 ${\bf 10}$ and
the ${\bf 16}$, while the other particles come only from the ${\bf 16}$.
This is because under SU(5)
 ${\bf 10}={\bf 5}+\bar{\bf 5}$
and it only contains states of gauge quantum numbers
of the $D^c$ and $L$.
The scalar masses are split according to
\be
m^2_{\tilde a}=m^2_{16}+\sin^2\theta(m^2_{10}-m^2_{16})\, ,
\label{splitsb}
\ee
which
 preserve SU(5) invariance
 because the $\langle {\bf 16}_H\rangle$ does not break this subgroup of
SO(10).
The fermion masses in this model are proportional to
$c_a\equiv[\cos\theta]_a$, the $\cos\theta$ of the field $a$.
For the third family, one has
\be
\frac{h_t}{h_b}=\frac{c_{t_R}c_{t_L}}{c_{b_R}c_{b_L}}
=\frac{1}{c_{b_R}}\, ,
\label{splitfb}
\ee
which for large values of $\rho$ ($M\ll \langle {\bf 16}_H\rangle$) leads to
\be
\frac{h_t}{h_b}\sim\rho\gg 1\, ;
\ee
this  accounts for the large mass difference $m_t\gg m_b$
without requiring a large ratio of the VEVs of the Higgs doublets.

The same mechanism of mass-splitting can be applied for the Higgs.
In the minimal SO(10) model the two light-Higgs doublets, $H_1$ and $H_2$,
are embedded in the ${\bf 10}_H$; their soft masses
are equal at $M_G$   and
 a severe fine-tuning is required to get the
correct
electroweak symmetry breaking \cite{hrs}.
Let us  introduce  to the minimal model
an extra ${\bf 16}_H$ and two $\overline{{\bf 16}}$s; one of them
 gets a VEV of $\calo(M_G)$:
\be
W=\overline{{\bf 16}}\left[M {\bf 16}_H+\lambda\, \overline{{\bf 16}}_H\,
{\bf 10}_H\right]+h\, {\bf 16}{\bf 10}_H\,  {\bf 16}\, .
\ee
In this model, one obtains
\bea
(m^2_{H_1}-m^2_{H_2})/m^2_{H_2}&=&\sin^2\theta(m^2_{16}-m^2_{10})/m^2_{10}\,
 ,\nonumber\\
h_t/h_b&=&1/\cos\theta\, ,
\eea
where $\sin\theta={\rho}/{\sqrt{1+\rho^2}}$ and $\rho=
\lambda\langle \overline{{\bf 16}}_H\rangle/M$.
Since the $M_P$--$M_G$ evolution \cite{pp}
leads to $m^2_{16}> m^2_{10}$, one finds the following
 nice correlation at $M_G$:
\be
h_t>h_b\Longleftrightarrow m^2_{H_1}>m^2_{H_2}\, ,
\ee
\ie,
$ m^2_{H_1}>m^2_{H_2}$, which  favours the electroweak breaking,
is related with the fact that $m_t>m_b$.

\end{document}